\documentclass{elsart}

\usepackage{graphicx}
\begin{document}

\begin{frontmatter}

\title{Complex dynamics in a one--block model for earthquakes}

\author{Ra\'ul Montagne}\ead{montagne@lftc.ufpe.br}\and\author{G. L. Vasconcelos\corauthref{glv}}
\corauth[glv]{Corresponding author. Tel.: +55-81-3271-8450; 
fax: +55-81-3271-0359}
\ead{giovani@lftc.ufpe.br}

\address{
Laborat\'orio de F\'{\i}sica Te\'orica e Computacional,
Departamento de F\'{\i}sica, Universidade Federal de Pernambuco,
50670-901, Recife, Brazil.
}

\begin{abstract}
A two-dimensional earthquake model that consists of a single block
resting upon a slowly moving rough surface and connected by two
springs to rigid supports is studied. Depending on the elastic
anisotropy and the friction force three generic regimes are possible:
i) pure creep; ii) pure stick-slip motion; and iii) a mixed regime.
In all cases the long-time dynamics (fixed point, periodic orbit or
chaos) is determined by the direction of the pulling velocity.  The
possible relevance of our findings to real faults is briefly
discussed.

\end{abstract}

\begin{keyword}
Earthquake models \sep Nonlinear dynamics \sep Creep \sep Stick-slip

\PACS 91.30.-f \sep 05.45.-a \sep 62.20.-x

\end{keyword}
\end{frontmatter}

\section{Introduction}

Since the seminal work of Burridge and Knopoff \cite{knopoff67},
spring-block models have been recognized as useful tools to study
earthquake dynamics \cite{turcotte_RPP}. Because they are simple to
treat, both theoretically and computationally, spring-block models can
provide important physical insights that might otherwise be much more
difficult to obtain.  Models with many blocks have been used to
investigate the origin of certain power laws that appear in the
statistics of earthquakes, such as the Gutenberg-Righter law for the
size-distribution of earthquakes \cite{turcotte_RPP}. Spring-block
models with only a few degrees of freedom, on the other hand, can
probe the basic dynamics of frictional sliding and are also of
interest in their own right as examples of low-dimensional nonlinear
dynamical systems.  For instance, two-block models have been found to
exhibit chaos \cite{turcotte,ryabov95,MSV99}, thus suggesting that
actual faulting might perhaps be a chaotic phenomena. Even in the
simplest case of a one-block model (with the block moving on a line)
the dynamics is highly nontrivial. Indeed, this system displays a
discontinuous transition from creep to stick-slip motion as one varies
a parameter governing the dependency of the friction force on the
velocity \cite{vasconcelos96}. A two-dimensional version of the
one-block model was recently considered \cite{ryabov01} and a richer
dynamics was found, including a chaotic regime where stick-slip events
occur intermittently amidst creep.

In this paper we present a detailed study of a 2D one-block model for
earthquakes, in which the block rests upon a moving planar surface and
is connected by springs to two perpendicular rigid walls, with a
velocity-weakening friction force acting between the slider and the
surface. An earlier discussion on this model was reported by Ryabov
and Ito \cite{ryabov01}. Here we give a much more complete analysis
and present several novel and surprising results. We show that in the
limit of very slow pulling the model is governed by three
dimensionless parameters, namely, the asymmetry parameter $\kappa$
corresponding to the ratio between the two spring constants, a
parameter $\gamma$ governing the decrease of friction 
with velocity, and the direction $\theta_\nu$ of the
pulling velocity. Depending on the values of $\kappa$
and $\gamma$, the model displays three possible generic regimes: i)
pure creep; ii) pure stick-slip motion; and iii) a mixed regime where
both creep and stick-slip occur. In each of these cases, the long-time
dynamics (fixed point, periodic orbit or chaos) is starkly dependent
on $\theta_\nu$, as we will see
below.

\section{The model}

\begin{figure}
\begin{center}
\includegraphics*[width=.55\columnwidth]{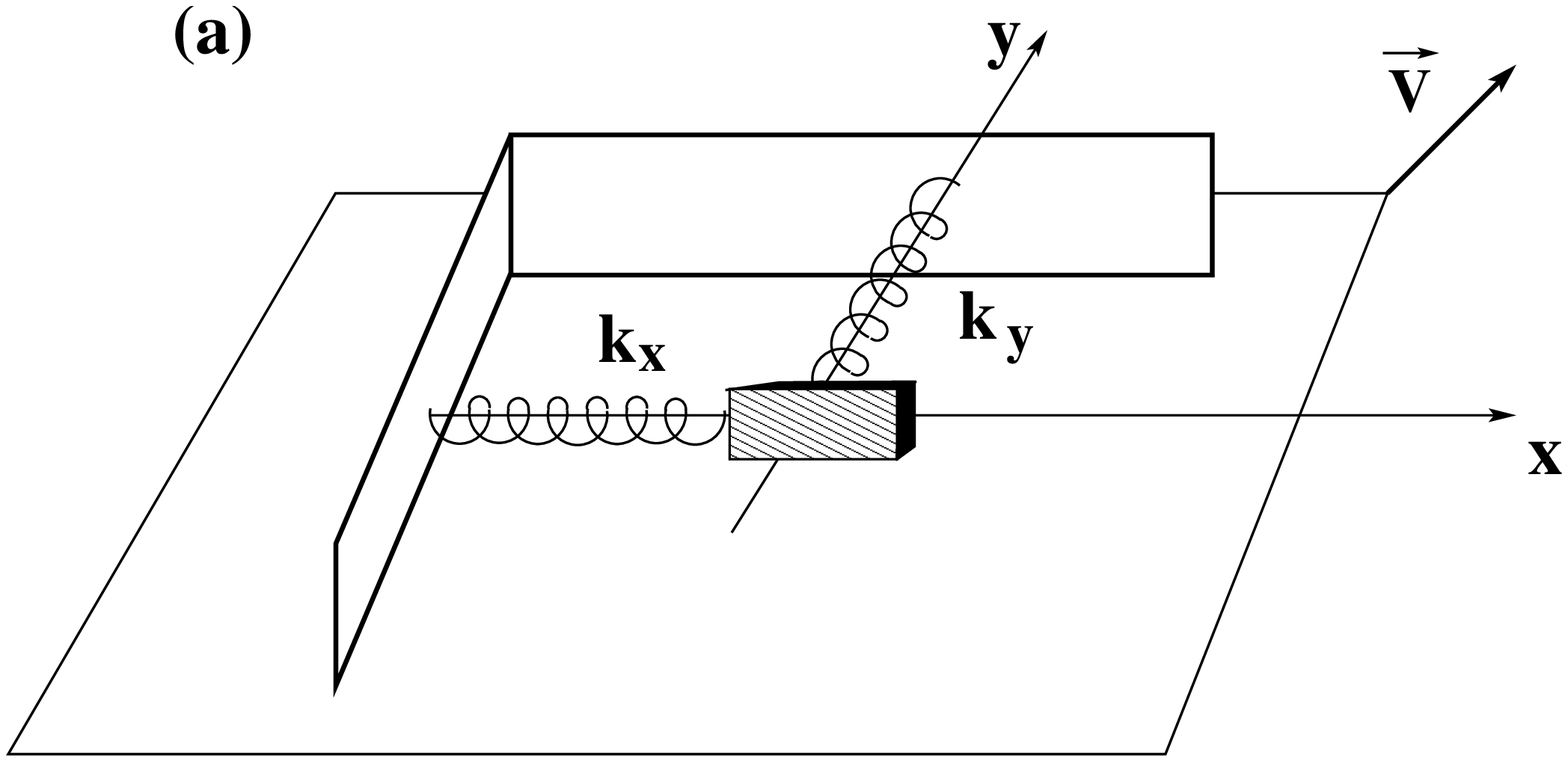}\hspace{0.7cm}
\includegraphics*[width=.34\columnwidth]{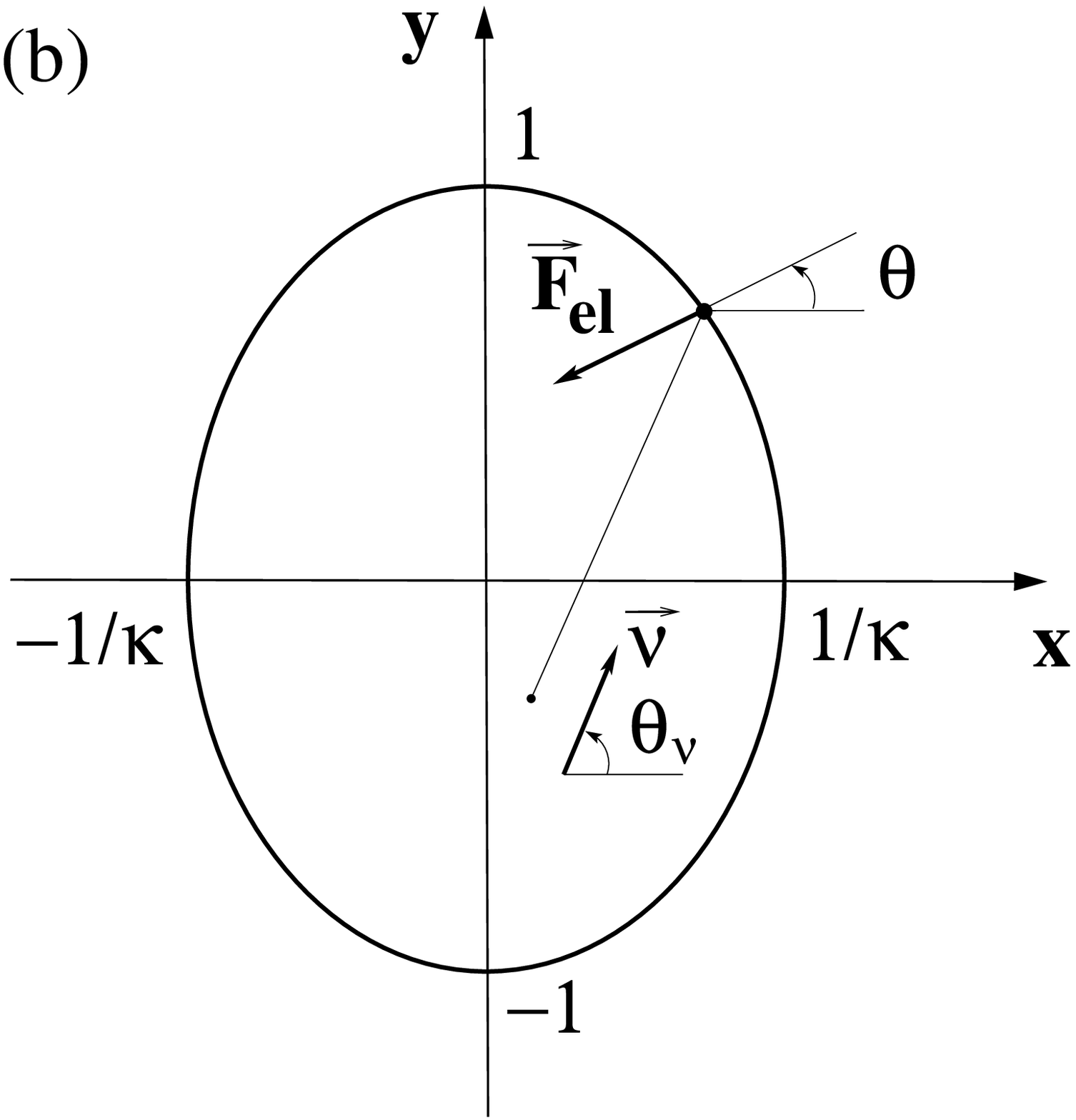}
\caption{(a) Two-dimensional spring--block model
and (b) the ellipse $C=\{(x,y)~|~\kappa^2 x^2+y^2=1\}$ where the
elastic force equals the static friction. Starting in a point
inside $C$, the block is dragged by the surface along the direction
$\theta_\nu$ of the pulling velocity until it hits a point on $C$,
whereupon slip begins. The angle $\theta$ 
denotes the direction opposite to the elastic force at this point.}
\label{modelinho}
\end{center}
\end{figure}

The model we consider consists of a block of mass $m$ connected to
motionless walls by two springs of stiffness $k_x$ and $k_y$, as shown
in Fig.~\ref{modelinho}a. The position of the block is described in a
system of coordinates $(x,y)$ fixed to the spring supports, with the
origin being placed at the point where the elastic force vanishes.
The block rests upon a surface that moves with a constant velocity
$\vec{V}$, where $|\vec{V}|\ll 1$.  There is friction between the
block and the moving surface, so that initially the block moves with
the substrate until the elastic force $\vec{F}_{\rm el}$ overcomes the
static friction force $F_0$, at which point the block slips with
respect to the surface.  In the slip phase of the motion, we adopt a
commonly used velocity-weakening friction law \cite{VVN}, in which the
magnitude of the friction force $\vec{F}_{\rm fr}$ is given by
$|\vec{F}_{\rm fr}| =F_0 \Phi(v_{\rm r}/V_{\rm f})$, where $v_{\rm r}=
\sqrt{(\dot{x} - V_x)^2+(\dot{y} - V_y)^2}$ is the relative velocity
of the slider with respect to the surface, $V_{\rm f}$ is a typical
velocity scale for the friction force, and $\Phi(x) = 1/(1+ x)$.
 
In the remainder of the paper we work with non-dimensional quantities,
where we have rescaled the length and time scales by $F_0/k_y$ and
$\sqrt{m/k_y}$, respectively, and velocities by $F_0/\sqrt{mk_y}$.  In
these units, the dimensionless parameters characterizing the problem
are the pulling velocity $\vec{\nu}$, the anisotropy parameter
$\kappa=k_x/k_y$, and the friction parameter $\gamma=1/2\nu_{\rm f}$,
where $\nu_{\rm f}$ is the dimensionless characteristic velocity of
the friction.  If we now denote by $C$ the curve of maximum
displacement, where the condition $|\vec{F}_{\rm el}|=F_0$ is
verified, it then follows that $C$ corresponds (in dimensionless
units) to the ellipse $\kappa^2x^2+y^2 = 1$; see Fig.~\ref{modelinho}b.
Starting from relative rest at a point inside $C$, the block will be
dragged by the surface until it reaches a point on $C$, upon which the
block starts to slide. In the slip phase the equations of motion are
given by
\begin{eqnarray}
\ddot{x} &=& -\kappa x + \Phi(2\gamma v_{\rm r})\cos\varphi_{\rm fr}, 
\label{eq:modela} \\
\ddot{y} &=&-y + \Phi(2\gamma v_{\rm r})\sin\varphi_{\rm fr}, 
\label{eq:model2}
\end{eqnarray}
where the angle $\varphi_{\rm fr}$ gives the direction of the friction
force. These equations  represent two coupled (nonlinear)
oscillators, and in this sense the system is somewhat similar to the
two-block model considered in Ref.~\cite{turcotte}. Our 2D
one-block model is, however, qualitatively richer in that it contains a new
parameter, namely, the direction of the pulling velocity which
strongly affects the dynamics of the system, as we will see shortly.

A final caveat about the model concerns the choice of direction of the
friction force. The problem comes about because at the onset of motion
the friction force is opposite to the elastic force, whereas during
the slip phase of the motion the friction force is expected to point
in the direction opposite to the relative velocity. Since the elastic
force is not in general collinear with the velocity, we should thus
prescribe a mechanism that allows the switching of direction of the
friction force.  Here we follow Ryabov and Ito \cite{ryabov01} and
assume that in the acceleration phase of the block motion, the
direction $\varphi_{fr}$ of the friction force is chosen according to
the following rule
\begin{equation}
 \varphi_{fr} = \pi+ \varphi_{el} -(\varphi_{el}-
 \varphi_{vr})[1-\exp(-\Omega v_{vr})] 
\label{friction}
\end{equation}
where $\varphi_{vr}$ and $\varphi_{el}$ denote the directions of the
relative velocity and the elastic force, respectively, and $\Omega$ is
a parameter describing the `memory' of the friction force.  (The
qualitative behavior of the model does not depend on the value of
$\Omega$ and so we have used $\Omega=50$ throughout the paper.) In the
deceleration phase, the friction force is kept opposite to the
velocity to ensure that the block always comes to a stop.

The dynamics of the model described above can be conveniently
described in terms of a one-dimensional map. To see this, let us first
parametrize the curve $C$ in terms of the angle $\theta$ defined by
the direction opposite to the elastic force with respect to the $x$
axis; see Fig.~\ref{modelinho}b. As a function of $\theta$, the
coordinates $(x_0,y_0)$ of a point on $C$ are given by the following
parametric equations: $x_0 =\kappa^{-1}\cos\theta, \, y_0
=\sin\theta$.  Now suppose that the block starts to slip from a given
point on $C$ labeled by $\theta$. The block will slide for a while
until it eventually comes to a stop with respect to the moving
surface. After this, the block sticks to the surface and is brought
back to another point $\theta'$ on $C$, where a new slip cycle begins,
and so on.  Thus, the overall dynamics of the model can be described
by a one-dimensional map $\theta'=f(\theta)$. Obviously the mapping
function $f(\theta)$ cannot be computed explicitly since this entails
solving (\ref{eq:modela}) and (\ref{eq:model2}) during the slip phase
of the motion. However, for a given set of parameters $\kappa$,
$\gamma$, and $\vec{\nu}$, the map $f(\theta)$ can be easily
constructed on the computer \cite{ryabov01}.

\section{Analysis of the model and results}

We begin the analysis of our model by considering the early stages of
a slip event. Suppose that the block, as it is being dragged by the
moving surface, reaches a point $(x_0,y_0)\in {C}$ where relative
motion begins.  At the onset of slip the block will move in the
direction of the elastic force $\vec{F}_{\rm el}=(-\kappa x_0,-y_0)$,
while the friction force will act in the opposite direction. In order
to study the nature of the ensuing motion, we linearize the equations
of motion in the directions parallel and perpendicular to the elastic
force. To accomplish this, it is convenient first to switch to the
reference frame where the substrate is at rest and then write the
equations of motion in the coordinates $(\xi,\eta)$, defined as the
block displacement in the directions parallel and perpendicular to
the elastic force, respectively. The resulting linearized equations are
\begin{eqnarray}
 \ddot{\xi} &=& -A\xi +\kappa(\kappa-1)x_0y_0
 \eta + 2\gamma \dot{\xi} + \alpha t, \label{eq:model3}\\
 \ddot{\eta} &=& \kappa(\kappa-1)x_0y_0 \xi - \kappa(\kappa
 x_0^2+y_0^2) \eta + \beta t,
\label{eq:model4}
\end{eqnarray}
where
\begin{equation}
A=\kappa^3 x_0^2+y_0^2, \quad 
 \alpha= -(\kappa^2 x_0\nu_x+y_0\nu_y) , \quad
 \beta=\kappa(y_0\nu_x-y_0\nu_y).
\label{eq:ab}
\end{equation}
We see from (\ref{eq:model4}) that in the limit $\nu\to 0$ 
the transversal
coordinate $\eta$ can grow only if so does the component $\xi$. We
can thus set $\eta=0$ in (\ref{eq:model3}), so that the linear
equation for the variable $\xi$ becomes
\begin{equation}
 \ddot{\xi} - 2\gamma \dot{\xi} + A\xi = \alpha t . \label{eq:lin}
\end{equation}

Equation (\ref{eq:lin}) for the case $A=1$ was studied in great detail
by one of the present authors \cite{vasconcelos96}.  There, it was
shown that in the limit that $\nu\to0$ the system undergoes a phase
transition at $\gamma=1$ in the following sense: for $\gamma<1$ there
is only creep, meaning that the linear approximation is always valid
and the amplitude of a slip event vanishes as $\nu\to0$, whereas for
$\gamma>1$ one has stick-slip motion in which case the slip amplitude
remains finite as $\nu\to0$.  These results can be trivially extended
to the general case given in (\ref{eq:lin}). Here the critical point
is reached when the condition $\gamma^2=A$ is satisfied, so that if
$\gamma^2<A$ we have creep, while for $\gamma^2>A$ stick-slip occurs.
Note, however, that since the parameter $A$ depends on the point
$(x_0,y_0)$ where the slip was initiated, it follows that for
a critical point to exist the ellipse defined by the critical
condition $A=\gamma^2$ must intersect the curve $C$.  If such an
intersection occurs, then the model displays both creep and
stick-slip, otherwise the motion is either pure creep or pure
stick-slip, as described next. 

\begin{figure}
\begin{center}
\includegraphics*[width=.48\columnwidth]{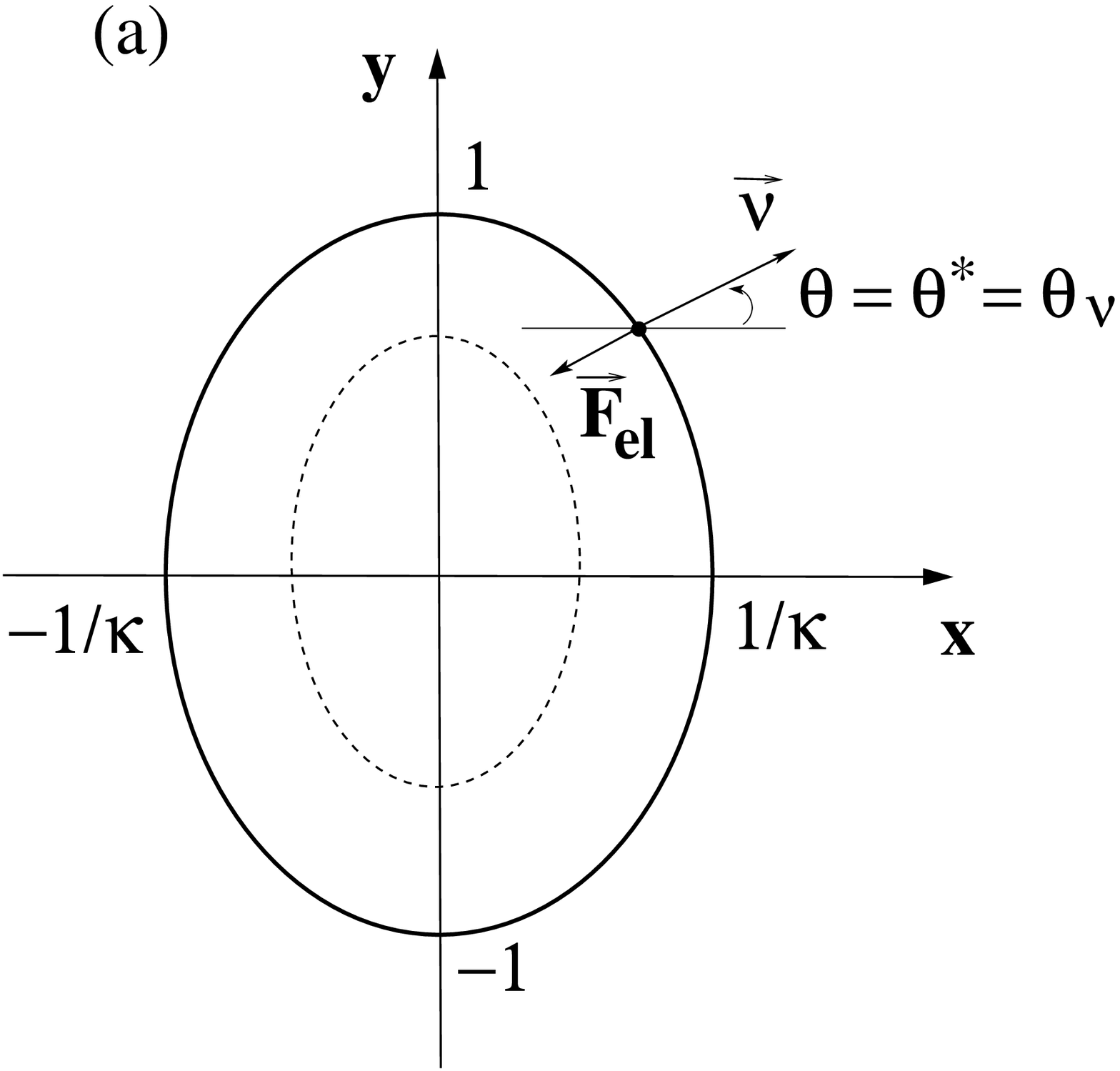}\hspace{0.5cm}\includegraphics*[width=.42\columnwidth]{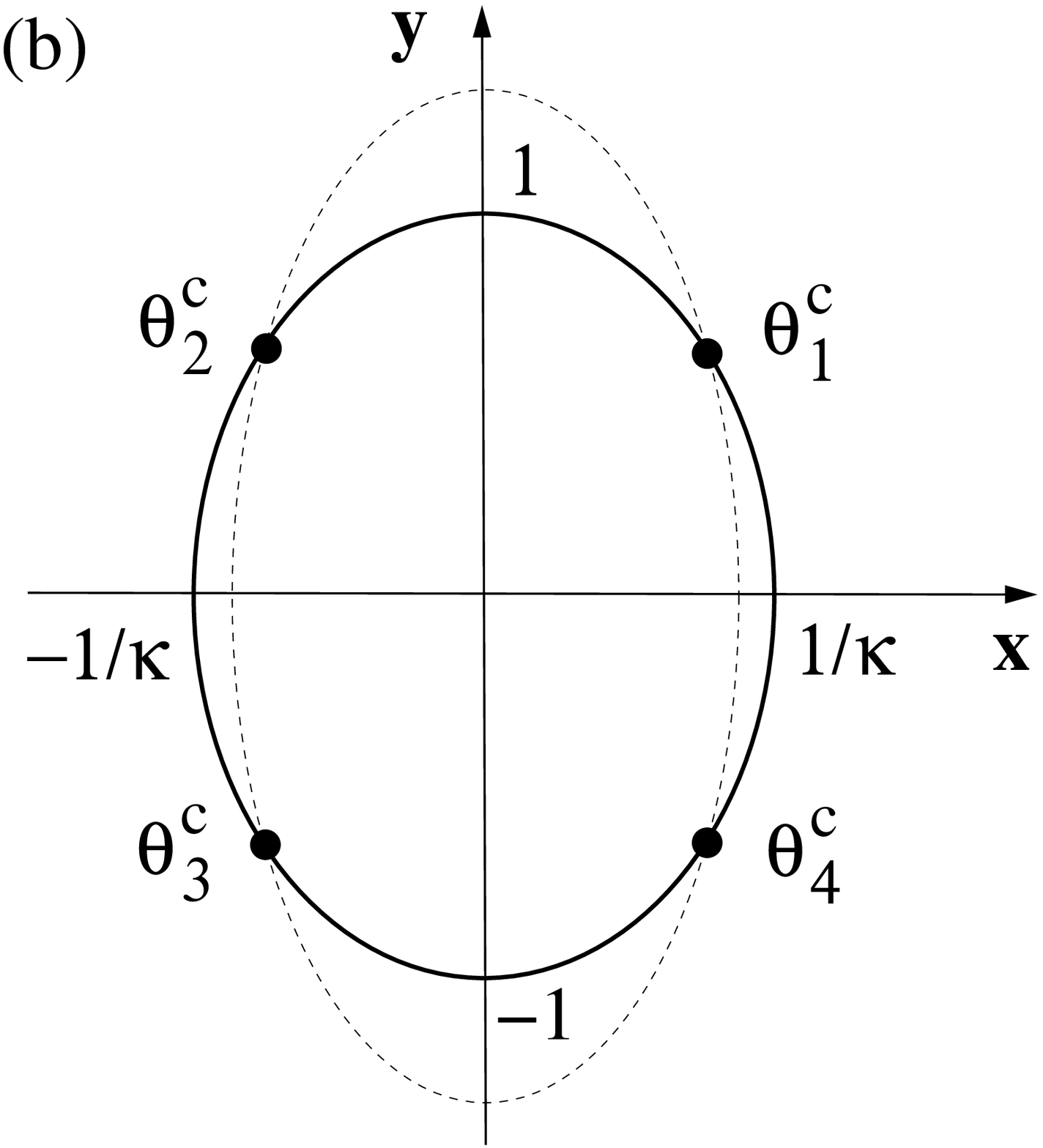}
\caption{The ellipse  $C$ of maximum displacement (solid line) 
and the critical ellipse (dashed line). We have $\gamma^2<1$ in (a)
and $1<\gamma^2<\kappa$ in (b).}
\label{ellipse}
\end{center}
\end{figure}

\noindent {\it Pure creep}: $\gamma < 1$. In this case the critical 
ellipse $A=\gamma^2$ is inside $C$ (Fig.~\ref{ellipse}a), so that
$\gamma^2 < A$ for any point $(x_0,y_0) \in C$, and hence the linear
approximation (\ref{eq:lin}) always holds (in the limit
$\nu\to0$). The long-time dynamics of the system is quite simple and
can be easily described in terms of the map $\theta'=f(\theta)$
introduced above.  Let us define by $\theta_\nu$ the direction of the
pulling velocity. Then for any initial condition the block will reach
the fixed point $\theta=\theta^*=\theta_\nu$, where the elastic force
is opposite to $\vec{\nu}$ (see Fig.~\ref{ellipse}a), so that the
resulting motion is essentially one-dimensional creep: starting from
the point on $C$ labeled by $\theta^*$, the block slides a very small
distance along the direction of the elastic force, stops, and then is
brought back to the same point $\theta^*$, and so on.

\begin{figure}
\begin{center}
\includegraphics*[width=.49\columnwidth]{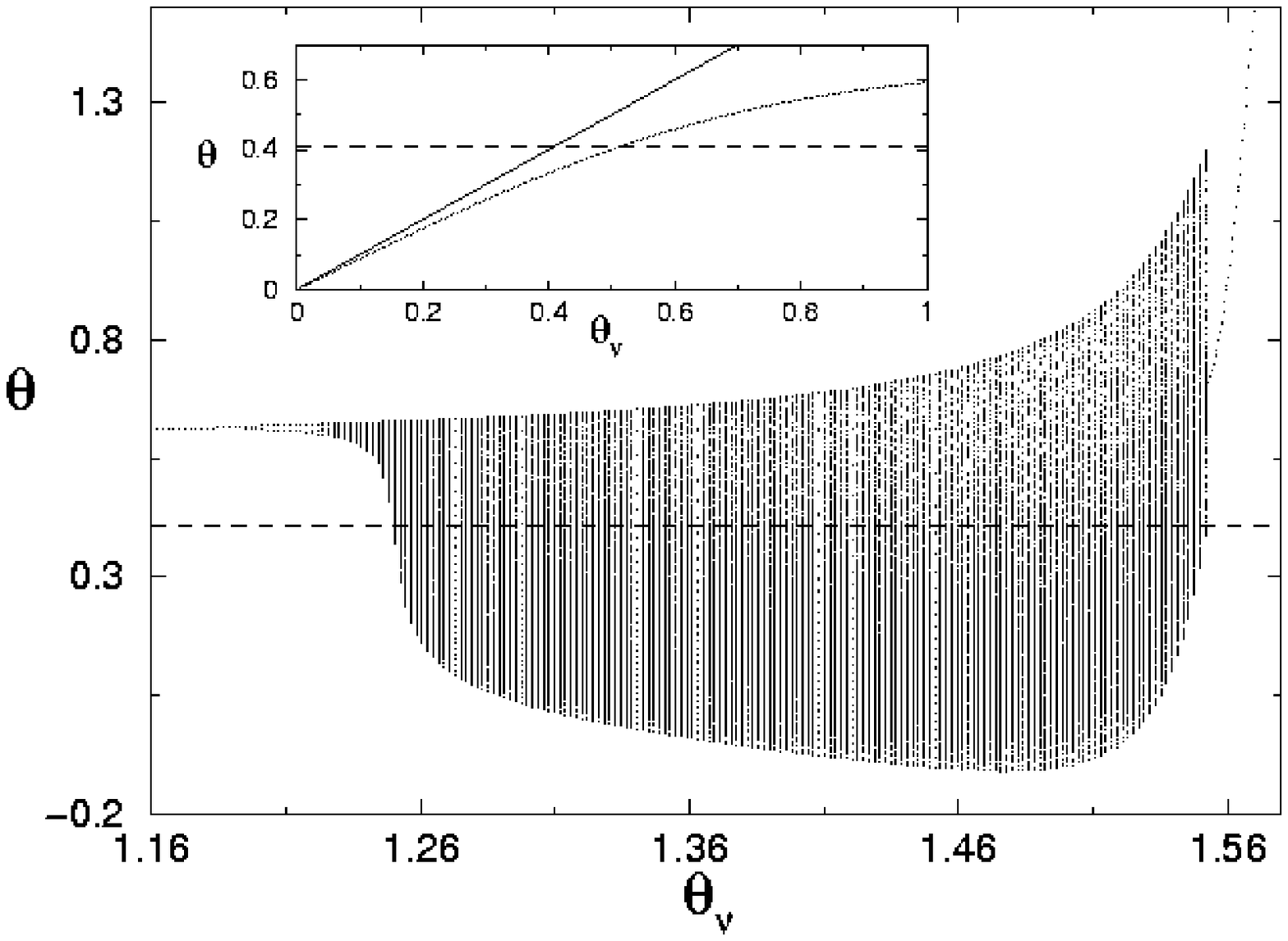}
\hspace{0.3cm}
\includegraphics*[width=.45\columnwidth]{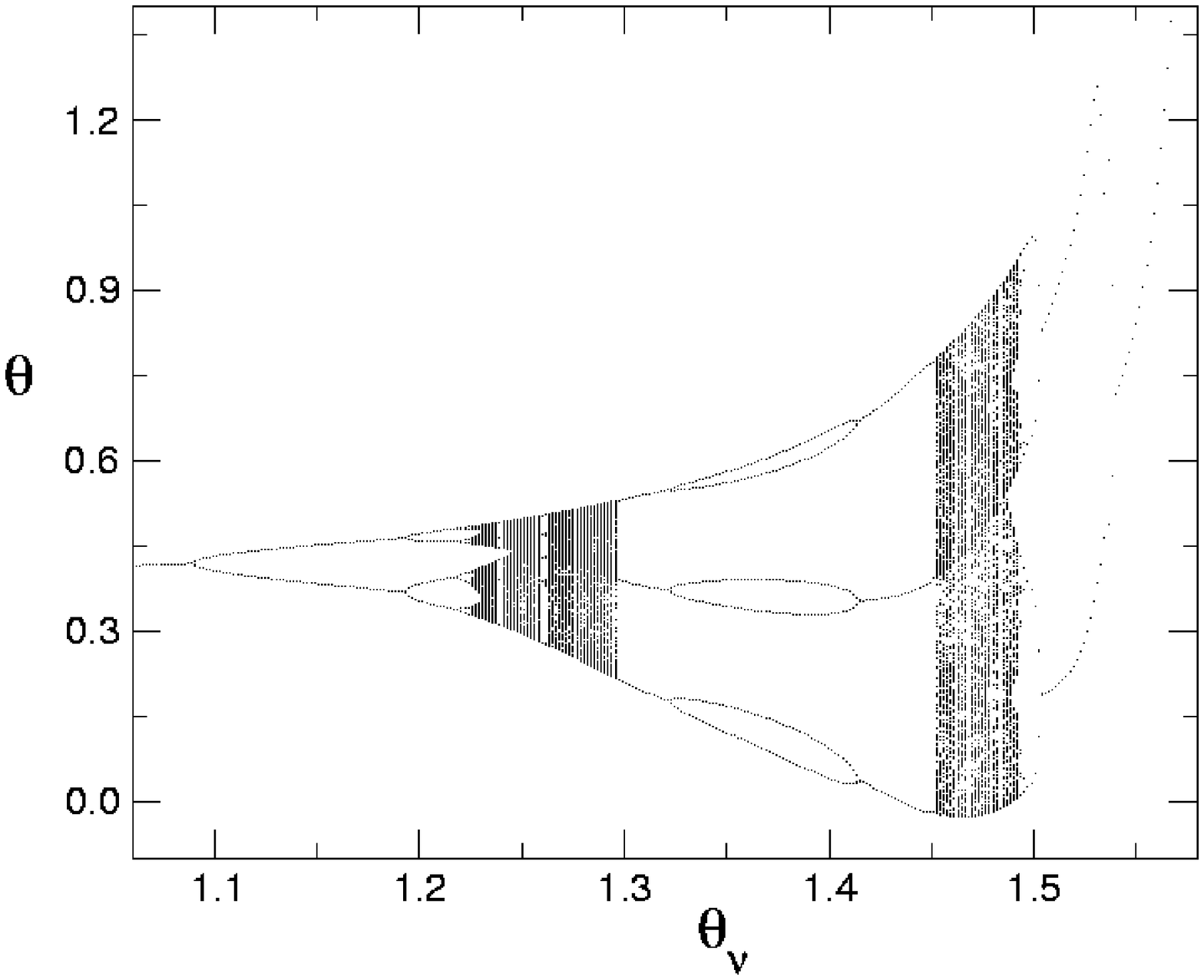}
\end{center}
\caption{Bifurcation diagram for the map  $f(\theta)$ in terms of  
the control parameter $\theta_\nu$. In (a) we have $\gamma=6.5$,
$\kappa=50$, $\nu=10^{-4}$, corresponding to the creep and stick--slip
regime, and the dashed line denotes the critical point
$\theta_c=0.41$. The inset shows the initial part of the bifurcation
diagram, with the solid straight line representing the $45^\circ$ line. In (b)
we have stick-slip only with $\gamma=6.5$, $\kappa=18.95$, and
$\nu=10^{-3}$.}
\label{fig2}
\end{figure}

\noindent {\it Creep and stick-slip}: $1< \gamma^2 < \kappa$. 
In this situation the critical ellipse and $C$ intersect each other at
four critical points $\theta_i^c$, where $\tan^2\theta_i^c
=\frac{\kappa-\gamma^2}{\gamma^2-1}$; see Fig.~\ref{ellipse}b.
Without loss of generality, let us assume that $\theta_\nu$ is
restricted to the first quadrant, so that for $0\le\theta<\theta_1^c$
we have creep ($\gamma^2<A)$, whereas for $\theta_1^c<\theta\le\pi/2$
stick-slip occurs ($\gamma^2>A$). Thus, for
$\theta_\nu\in[0,\theta_c)$ the creep fixed-point
$\theta=\theta^*=\theta_\nu$ is the only attractor. This is
illustrated in the inset of Fig.~\ref{fig2}a, where we plot the
initial part of the bifurcation diagram for the map $f(\theta)$ with
$\gamma=6.5$, $\kappa=50$, $\nu=10^{-4}$, and $\theta_\nu$ as the
control parameter. In this inset, the dashed line represents the
critical point $\theta_c=0.41$ and we see that for
$\theta_\nu<\theta_c$ the fixed point indeed follows closely the
$45^\circ$ line (solid straight line), with the small deviation being
caused by the finite value of $\nu$ that smoothes out the transition
from creep to stick-slip.  As $\theta_\nu$ increases past $\theta_c$,
the system remains in a stick-slip fixed point for awhile but
eventually undergoes a sequence of period-doubling bifurcations
leading to chaos, as seen in the main bifurcation diagram shown in
Fig.~\ref{fig2}a. For larger $\theta_\nu$, the system behaves
intermittently \cite{ryabov01} in the following sense: the block
spends a long time in the creep region (below the dashed line) until
it eventually crosses the critical point $\theta_c$, after which it
undergoes a large slip event and is reinjected back into the creep
region, and so on. Finally, as $\theta_\nu$ approaches $\pi/2$ there
is a reverse period-doubling cascade back to a fixed point.

\begin{figure}
\begin{center}
\includegraphics*[width=.5\columnwidth,angle=90]{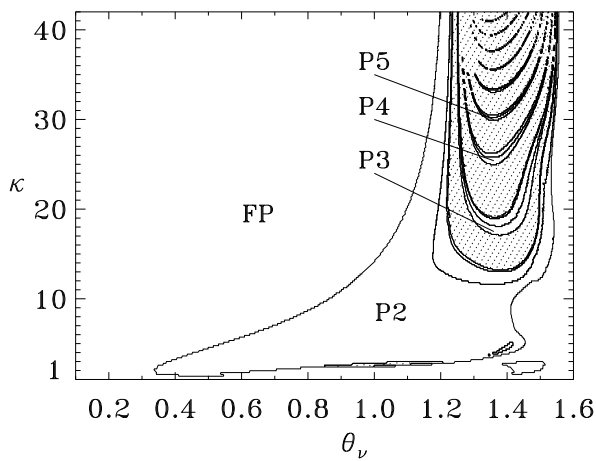}
\end{center}
\caption{Phase diagram in the plane 
$(\theta_\nu,\kappa)$ for the pure stick-slip regime.  $FP$ denotes 
the region with a stable fixed point, $Pn$ indicates regions of
period-$n$ orbits, and the shaded areas represent chaotic regions. Here
we used $\gamma=6.5$ and $\nu=10^{-3}$.}
\label{fig:contour}
\end{figure}

\noindent {\it Pure stick-slip}: $\gamma^2 > \kappa$. Here
the critical ellipse is exterior to $C$ and so we have $\gamma^2>A$
for any point on $C$, hence the block always undergoes stick-slip
motion.  As the parameters $\kappa$ and $\theta_\nu$ are varied (with
$\gamma$ fixed), the system now displays a very rich behavior that is
summarized in the {\it phase diagram} shown in Fig.~\ref{fig:contour},
whose main features we shall briefly discuss.  First we see that for
small values of $\theta_\nu$ the system always reaches a stable fixed
point (region labeled by $FP$), where slip events occur
periodically. As $\theta_\nu$ increases (with $\kappa$ fixed), what
follows next depends on the value of $\kappa$. For $\kappa$ very close
to 1 the fixed point remains stable for all values of $\theta_\nu$;
see bottom of Fig.~\ref{fig:contour}.  However, for larger $\kappa$
the fixed point eventually goes unstable and a orbit of period two is
born (region $P2$). The fate of this period-2 orbit, as $\theta_\nu$
increases further, also depends on the value of $\kappa$. For small
$\kappa$, e.g., $\kappa=10$, the system simply undergoes a reverse
period-doubling bifurcation back to the fixed point. For intermediate
values of $\kappa$, say, $\kappa=16$, one observes a full
period-doubling cascade leading to chaos (shaded region), followed by
a reverse cascade.  For larger $\kappa$, windows of periodic orbits
and their associated cascades appear inside the chaotic
region. Several of these windows are clearly visible in
Fig.~\ref{fig:contour}, although we have labeled only the first three
of them, namely, the windows of period 3, 4, and 5.  Note also that
depending on the value of $\kappa$ the secondary period-doubling
cascades may not develop fully.  An example of this is given in
Fig.~\ref{fig2}b where we show the bifurcation diagram for the map
$f(\theta)$ with $\gamma=6.5$ and $\kappa=18.95$.

\section{Conclusions}

In conclusion, we have studied a two-dimensional one-block model for
earthquakes that shows a very rich behavior and is able to reproduce a
host of relevant dynamical regimes such as creep, stick-slip (periodic
and chaotic) and intermittent stick-slip amidst creep. Particularly
interesting is our finding that the dynamics of the model is strongly
dependent on the direction of the pulling velocity. This property thus
suggests that, in addition to the friction force acting on a fault and
the stiffness of the loading system, the direction of shear might also
play an important role in determining the fault seismic activity (or lack
thereof). Such mechanism could perhaps help explaining, for instance,
why different segments of a fault system may exhibit distinct seismic
patterns although they presumably have similar geophysical
properties. These are interesting possibilities that certainly deserve
to be investigated further.

\begin{ack} We thank M.~A.~F.~Gomes for a critical reading 
of the manuscript. Financial support from the Brazilian agencies CNPq
and FINEP and from the special research programs PRONEX and CTPETRO is
acknowledged.  R.~M.~acknowledges financial support from
C.S.I.C.~(Uruguay) and the Programa de Desarrollo de Ciencias
B\'asicas (PEDECIBA, Uruguay).
\end{ack}


\end{document}